\documentclass[twocolumn,showpacs,preprintnumbers,nofootinbib,prd,superscriptaddress,10pt,aps]{revtex4-1}

\usepackage{graphicx,amssymb,amsmath,amsthm,amsfonts,epsfig,epsf,epstopdf,times}
\usepackage[linktocpage]{hyperref}
\usepackage[usenames]{color}
\usepackage{epstopdf}
\usepackage{multirow}
\usepackage{bm}
\usepackage{dcolumn}
\usepackage[latin1]{inputenc}
\usepackage{latexsym}
\usepackage{rotating}
\usepackage{hyperref}
\usepackage{color}
\usepackage{longtable}
\usepackage{enumerate}
\usepackage{tensor}
\usepackage{url}
\definecolor{coolblack}{rgb}{0.0, 0.18, 0.39}
\definecolor{darkred}{rgb}{0.5,0,0}
\definecolor{darkgreen}{rgb}{0,0.5,0}
\definecolor{darkblue}{rgb}{0,0,0.5}
\definecolor{lapislazuli}{rgb}{0.15, 0.38, 0.61}
\definecolor{venetianred}{rgb}{0.78, 0.03, 0.08}
\definecolor{bleudefrance}{rgb}{0.19, 0.55, 0.91}
\definecolor{dogwoodrose}{rgb}{0.84, 0.09, 0.41}
\definecolor{dogwoodrose}{rgb}{0.84, 0.09, 0.41}
\hypersetup{colorlinks=true, citecolor=darkblue, linkcolor=darkblue, 
urlcolor = darkblue}

\def\pa{\partial}

\def\nn{\nonumber}

\newcommand{\ben}{\begin{enumerate}}
\newcommand{\een}{\end{enumerate}}

\def\be{\begin{equation}}
\def\ee{\end{equation}}
\def\bea{\begin{eqnarray}}
\def\eea{\end{eqnarray}}
\def\nn{\nonumber}
\newcommand{\beq}{\begin{eqnarray}}
\newcommand{\eeq}{\end{eqnarray}} 
\newcommand{\ba}{\begin{align}}
\newcommand{\ea}{\end{align}}
\newcommand{\op}{\left}
\newcommand{\cl}{\right}

\begin{document}

\title{Perturbed black holes in Einstein-dilaton-Gauss-Bonnet gravity:\\ Stability, ringdown, and gravitational-wave emission}

\author{Jose Luis Bl\'azquez-Salcedo}
\affiliation{Department of Physics, University of Oldenburg, Oldenburg, 26111, Germany}

\author{Caio F. B. Macedo}
\affiliation{CENTRA, Departamento de F\'{\i}sica, Instituto Superior T\'ecnico, Universidade de Lisboa, Avenida~Rovisco Pais 1, 1049 Lisboa, Portugal}

\author{Vitor Cardoso}
\affiliation{CENTRA, Departamento de F\'{\i}sica, Instituto Superior T\'ecnico, Universidade de Lisboa, Avenida~Rovisco Pais 1, 1049 Lisboa, Portugal}
\affiliation{Perimeter Institute for Theoretical Physics, 31 Caroline Street North
Waterloo, Ontario N2L 2Y5, Canada}
\affiliation{Theoretical Physics Department, CERN, CH-1211 Gen\`eve 23, Switzerland}

\author{Valeria Ferrari}
\affiliation{Dipartimento di Fisica, ``Sapienza'' Universit\`a di Roma \& Sezione INFN Roma1, Piazzale Aldo Moro 5, 00185, Roma, Italy}

\author{Leonardo Gualtieri}
\affiliation{Dipartimento di Fisica, ``Sapienza'' Universit\`a di Roma \& Sezione INFN Roma1, Piazzale Aldo Moro 5, 00185, Roma, Italy}

\author{Fech Scen Khoo}
\affiliation{Department of Physics and Earth Sciences, Jacobs University, Bremen, 28759, Germany}

\author{Jutta Kunz}
\affiliation{Department of Physics, University of Oldenburg, Oldenburg, 26111, Germany}

\author{Paolo Pani}
\affiliation{Dipartimento di Fisica, ``Sapienza'' Universit\`a di Roma \& Sezione INFN Roma1, Piazzale Aldo Moro 5, 00185, Roma, Italy}
\affiliation{CENTRA, Departamento de F\'{\i}sica, Instituto Superior T\'ecnico, Universidade de Lisboa, Avenida~Rovisco Pais 1, 1049 Lisboa, Portugal}

\begin{abstract}

Gravitational waves emitted by distorted black holes---such as those arising from the coalescence of two neutron stars or black holes---carry not only information about the corresponding spacetime but also about the underlying theory of gravity.
Although general relativity remains the simplest, most elegant and viable theory of gravitation,
there are generic and robust arguments indicating that it is not the ultimate description of the gravitational universe. 
Here, we focus on a particularly appealing extension of general relativity, which corrects Einstein's theory through the addition of terms which are second order in curvature: the topological Gauss-Bonnet invariant coupled to a dilaton.
We study gravitational-wave emission from black holes in this theory and {\bf(i)} find strong evidence that black holes are linearly (mode) stable against both axial and polar perturbations,
{\bf(ii)} discuss how the quasinormal modes of black holes can be excited during collisions involving black holes, and finally
{\bf(iii)} show that future ringdown detections with a large signal-to-noise ratio would improve current constraints on the coupling parameter of the theory.
\end{abstract}

\maketitle


\section{Introduction}
\label{sec:intro}
The historical detection of gravitational waves (GWs) by the LIGO/Virgo Collaboration has marked the beginning of a new era in astrophysics
and the birth of GW astronomy~\cite{Abbott:2016blz}. The next generation of detectors will routinely observe the coalescence
of compact objects, such as black holes (BHs) and neutron stars. These observations will probe, for the first time, the highly dynamical
regime of strong-field gravity and may provide the answer to long-standing
issues~\cite{Cardoso:2016rao,Nakano:2016sgf,Yunes:2016jcc}. Is cosmic censorship
preserved in violent gravitational interactions? Do GW observations carry incontrovertible evidence for the event horizon of BHs? Can we pinpoint, in gravitational waveforms, the signature of the light ring or of ergosurfaces?

Simultaneously, the entire coalescence process can be used to constrain gravity theories in novel ways~\cite{Berti:2015itd,TheLIGOScientific:2016src,Yunes:2016jcc}. That general relativity (GR) is not the ultimate theory of gravity is a possibility that should be entertained
in the light of several observations (such as those related to the dark-matter and the dark-energy problems),
and of the difficulty to reconcile GR with quantum field theory~\cite{Berti:2015itd}.
Although such an extension of GR is unknown---and a robust spacetime parametrization in strong-field gravity is lacking---GW observations will help us to exclude or to strongly constrain
wide classes of alternative theories. The inspiral stage, for example, when the two objects are far apart, can teach us about possible extra radiation channels~\cite{TheLIGOScientific:2016src,Barausse:2016eii,Cardoso:2016olt,Yunes:2016jcc}, while the final ringdown stage---when the end-product is relaxing to its final state---provides for remarkable tests of GR, through the measurement of the characteristic quasinormal modes (QNMs)~\cite{Berti:2009kk}.
In GR, as well as in essentially any relativistic theory of gravity, BHs are extremely simple objects described by only a handful of parameters. Accordingly, their QNMs are completely characterized
by only a few parameters as well. For example, Kerr BHs in GR are characterized by their mass and angular momentum, and so are their QNMs.
In a nutshell, measurement of one single QNM (i.e, a ringing frequency and a decay time scale~\cite{Berti:2005ys,Berti:2009kk}) allows for a determination of the BH mass and angular momentum. The measurement of a second QNM tests GR~\cite{Berti:2005ys,Berti:2007zu,Gossan:2011ha,Meidam:2014jpa}. In the context of modified theories of gravity, a second QNM can be used to measure possible extra coupling parameters, as was shown recently
for a theory with an extra vector degree of freedom~\cite{Cardoso:2016olt}.

Some of the most viable and appealing modifications of gravity are those obtained via the inclusion of extra scalar fields---such as scalar-tensor theories of gravity---or of higher-curvature terms in the action, or both. Higher-order gravity is generically motivated by UV corrections, which also arise naturally in some low-energy truncations of string theories. The paradigmatic case, and the one we focus on here, is Einstein-dilaton-Gauss-Bonnet (EDGB) gravity, described by the action~\cite{Kanti:1995vq,Berti:2015itd}
\be
S=\int d^4 x\frac{\sqrt{-g}}{16\pi}\op(R-\frac{1}{2}\pa_a\phi\pa^a\phi+\frac{\alpha}{4}e^{\phi}{\cal R}_{\rm GB}^2\cl)+S_{m}\,,
\label{eq:action}
\ee
where
\be
{\cal R}_{\rm GB}^2={R_{abcd}} R^{abcd}-4 {R_{ab}} R^{ab}+R^2\,,
\ee
is the Gauss-Bonnet topological term, $S_m$ represents the matter sector and we use (throughout this work) units for which the Newton's constant and the speed of light are unity, $G=c=1$.
Current best constraints on the coupling constant $\alpha$ are $\sqrt{\alpha}<10\,{\rm km}$~\cite{Yagi:2012gp,Berti:2015itd}\footnote{We note a typo in the review~\cite{Berti:2015itd}. In the notation used in the review, Eq.~(2.26) should read $\sqrt{|\alpha_{\rm GB}|}\lesssim 5\times10^5\,{\rm cm}$.}.

We will close two important gaps in the literature concerning BHs in this theory:
we first find strong evidence that static EDGB BHs are linearly stable,
and then compute gravitational waveforms from plunging particles,
which can be argued to be an indicator of how BH collisions proceed in this theory.
Finally, we discuss the constraints on the coupling parameter $\alpha$ from current and future ringdown observations.

\section{Framework}
\label{sec:review}
The equations of motion obtained by extremizing \eqref{eq:action} with respect to the metric and dilaton field are given by~\cite{Kanti:1995vq}
\begin{align}
\Box \phi&=\frac{\alpha}{4}e^{\phi}{\cal R}_{\rm GB}^2\label{eq:scalarphi}\,,\\
G_{ab}&=\frac{1}{2}\pa_a\phi\pa_b\phi-\frac{1}{4}g_{ab}
(\pa_c\phi) (\pa^c\phi)
-\alpha {\cal K}_{ab}+8\pi T_{ab} \label{eq:einsteineq}\,,
\end{align}
where $G_{ab}=R_{ab}-\frac{1}{2}g_{ab}R$ is the Einstein tensor, $T_{ab}$ is the matter stress-energy tensor,
and 
\be
{\cal K}_{ab}=(g_{ac}g_{bd}+g_{ad}g_{bc})\epsilon^{idjk}\nabla_{l}\op(\tilde{R}^{cl}_{~~jk}\pa_i e^{\phi}\cl)\,,
\ee
where $\epsilon^{abcd}$ is the contravariant Levi-Civita tensor, $\tilde{R}^{ab}_{~~cd}=\epsilon^{abij}R_{ijcd}$.
\subsection{BH solutions in EDGB gravity}

BHs in EDGB gravity are scalar-vacuum solutions of the above equations, which were first constructed analytically in spherical symmetry, in the small-coupling regime~\cite{Mignemi:1992nt},
\be
\zeta:=\frac{\alpha}{M^2}\ll1 \,,
\ee
where $M$ is the BH mass. In spherical symmetry, the line element reads
\be
ds^2=-A(r)dt^2+B(r)^{-1}dr^2+r^2d\Omega^2\,,
\label{eq:linel}
\ee
where $d\Omega^2$ is the standard unit 2-sphere line element. 
Both functions $A(r)$ and $B(r)$ and the scalar field $\phi$ can be expanded in powers of the coupling parameter $\zeta$, and the corresponding solutions can be found by solving the field equations~\eqref{eq:scalarphi}--\eqref{eq:einsteineq} perturbatively (see also Ref.~\cite{Yunes:2011we}).
Further details are given in Appendix \ref{app:motion}.

Nonperturbative solutions were investigated numerically in Ref.~\cite{Kanti:1995vq} for static geometries and in Ref.~\cite{Pani:2009wy} for slowly rotating BHs to first order in the spin.
It was shown that static BH solutions exist only up to a maximum value of $\zeta$, namely~\cite{Pani:2009wy}
\be
0\leq \zeta \lesssim 0.691\,.
\ee
Because $\zeta$ is strictly less than unity, higher-order perturbative expansions~\cite{Maselli:2015tta} are
accurate almost in the entire parameter space.

Slowly rotating solutions were described numerically in Ref.~\cite{Pani:2009wy}, and analytically in the small-coupling regime in Refs.~\cite{Pani:2011gy,Ayzenberg:2014aka}, and were recently extended to higher order in the coupling and in the spin parameter~\cite{Maselli:2015tta}. In the latter case, the line element 
(as well as the scalar field) can be expanded in a complete basis of orthogonal functions according to their symmetry properties~\cite{Maselli:2015tta}.

Numerical solutions describing rotating BHs for arbitrary coupling and spin were found in Ref.~\cite{Kleihaus:2011tg} and have been recently thoroughly discussed in Ref.~\cite{Kleihaus:2015aje}.

The linearized mode stability of spherically symmetric BHs against radial fluctuations was studied in Ref.~\cite{Kanti:1997br}, whereas axial gravitational perturbations were studied in Ref.~\cite{Pani:2009wy}. In the polar sector, the linear stability of EDGB BHs was analyzed in Ref.~\cite{Ayzenberg:2013wua}, focusing in the particular regimes where perturbations are dominantly gravitational or scalar, as well using a high-frequency analysis of the perturbations.
In addition, axial quasinormal modes of neutron stars in EDGB theory were studied in Ref.~\cite{Blazquez-Salcedo:2015ets}.

Due to the cumbersome field equations, there is currently no result concerning the stability of nonrotating EDGB BHs for the most relevant gravitational polar sector, and there is no stability analysis for rotating solutions. 
In this work we partly fill this gap by performing a full linear (mode) stability analysis of static EDGB BHs.

\subsection{Perturbed BHs in EDGB}\label{subsec:pert}
We are interested in understanding how BHs in EDGB theory respond to small perturbations, as those induced by a
fluctuation of the metric or of the dilaton or by a small external perturbing object. We focus our attention in both
dilaton fluctuations in vacuum and those induced by a small pointlike particle plunging into the BH. The first case will describe the late-time behavior of perturbed EDGB
BHs, which also dominates the ringdown signal from a distorted BH formed in a coalescence. Pointlike particles, on the
other hand, are a good proxy for small BHs or neutron stars falling into massive BHs, but are also known to provide
reasonably accurate estimates even for equal-mass BH
collisions~\cite{Berti:2010ce,Sperhake:2011ik,Tiec:2014lba}.

Pointlike particles of mass $\mu\ll M$ are modeled by~\cite{Zerilli:1971wd,Poisson:2011nh,Breuer:1974uc}
\begin{align}
S_m=\mu \int d\tau\sqrt{-g}\,,
\end{align}
with $d\tau=d\lambda \sqrt{-g_{ab}\dot x^a\dot x^b}$. We will consider pointlike objects with a scalar charge which is entirely due to the Gauss-Bonnet coupling. In other words, we will investigate BH spacetimes of which the scalar charge arises purely from the Gauss-Bonnet term. The theory we consider contemplates no other couplings to matter, and therefore
pointlike particles follow geodesics. This need not be the case generically, and nontrivial couplings to matter can be envisioned~\cite{Quinn:2000wa,Poisson:2011nh}. These couplings will certainly influence the motion of particles and the radiation in collisions but will not affect the intrinsic ringdown properties of the spacetime.

We consider a spherically symmetric EDGB BH distorted by either the pointlike particle or through some fluctuation in the metric or scalar field. At the linearized level, the full geometry is described by
\begin{align}
g_{ab}&=g_{ab}^{(0)}+\varepsilon\, h_{ab}\,,\\
\phi&=\phi_0(r)+\varepsilon\,\delta\phi\,,
\end{align}
where $\varepsilon \ll 1$ is a bookkeeping parameter, $g_{ab}^{(0)}$ is described by \eqref{eq:linel} and $\phi_0(r)$ is the corresponding background scalar. As background solution, we consider both a perturbative solution~\cite{Maselli:2015tta} up to ${\cal O}(\zeta^6)$, and a numerical solution for arbitrary values of $\zeta$.

The fluctuations $h_{ab}$ and $\delta \phi$ are functions of $(t,r,\theta,\varphi)$. Einstein's equations can be further simplified by Fourier transforming these  quantities
and by expanding them in (tensor and scalar) spherical harmonics, e.g.,
\be
\delta\phi(t,r)=\frac{1}{\sqrt{2\pi}}\int d\omega \frac{\phi_1(\omega,r)}{r}Y^{lm} e^{-i\omega t}\,,\label{Fourier}
\ee
where $Y^{lm}$ are the standard spherical harmonics and $\omega$ is a Fourier frequency.

The metric perturbations can be decomposed in terms of tensorial spherical harmonics~\cite{Regge:1957td,Zerilli:1971wd}. By using this decomposition, perturbations naturally split into two sectors according to their parity (either axial or polar). Axial perturbations of EDGB BHs are simpler, because they are decoupled from the scalar-field perturbations~\cite{Pani:2009wy}. Here, we shall consider the two sectors of the perturbations.

In the Regge-Wheeler gauge~\cite{Regge:1957td}, the polar sector of the metric perturbations is given by 
\begin{align}
&h_{ab}=\op[
\begin{array}{cccc}
A H_0 &  H_1  & 0 & 0 \\
H_1 & H_2/B & 0 & 0\\
0 & 0 & r^2 K & 0\\
0 & 0 & 0 & r^2 \sin^2\theta K
\end{array}\cl] Y^{lm} \,,
\label{hab13}
\end{align}
while the axial sector reads
\begin{align}
&h_{ab}=\op[
\begin{array}{cccc}
0                         & 0                         & 0 & \sin\theta~ h_0\pa_\theta\\
0                         & 0                         & 0 & \sin\theta~ h_1\pa_\theta\\
0                         & 0                         & 0 & 0\\
\sin\theta~ h_0\pa_\theta & \sin\theta~ h_1\pa_\theta & 0 & 0
\end{array}\cl]\,Y^{lm} \,.
\end{align}
In the above definitions, the metric perturbations depend only on $t$ and $r$. Note that we have already specialized the spacetime to axial symmetry: for the cases handled here, one can always rotate the coordinate axis such
that the spacetime is axially symmetric.
We shall Fourier-decompose these perturbation functions 
as, e.g., $X(t,r)=(2\pi)^{-1/2}\int d\omega X(\omega,r)e^{-i\omega t}$. 

Likewise, the stress-energy tensor can be decomposed into spherical harmonics~\cite{Zerilli:1971wd,Sago:2002fe}. In the radial plunging case considered here, the particle only disturbs the spacetime in the polar sector, and its stress-energy tensor can be written as
\be
T_{ab}=
\left[
\begin{array}{cccc}
A_{lm}^{(0)} & \frac{i}{\sqrt{2}}A_{lm}^{(1)} & 0 & 0 \\
 \frac{i }{\sqrt{2}}A_{lm}^{(1)} & A_{lm}& 0 & 0 \\
 0 & 0 & 0 & 0 \\
 0 & 0 & 0 & 0 \\
\end{array}
\right]Y^{lm}\,,
\ee
where the functions $A_{lm}^{(0)}$, $A_{lm}^{(1)}$ and $A_{lm}$, like the dilaton and metric perturbations, can be Fourier decomposed, such that they depend only on $r$ and $\omega$. The explicit form of these functions is given in Appendix~\ref{app:line}.

The linearized dynamics is governed by a coupled system which can be obtained by expanding the field equations \eqref{eq:scalarphi}--\eqref{eq:einsteineq} up to first order in the perturbation functions $h_{ab}$ and $\delta\phi$. The equations take the schematic form (no sum on $j$) 
\be
\frac{d}{dr}{\bm{\Psi}}_j+\bm{V}_{j}\bm{\Psi}_j=\bm{S}_j\,,
\label{eq:perturbations}
\ee
where $j=({\rm p},{\rm a})$ for the polar and axial sectors, respectively, $\bm{\Psi}_j$ is a column vector, the components of which are $\bm{\Psi}_{\rm p}\equiv (H_1,K,\phi_1,\phi_1')$ for the polar sector and $\bm{\Psi}_{\rm a}\equiv (h_1,h_0)$ for the axial sector, respectively. The matrices $\bm{V}_{j}$ describe the coupling among the perturbations and depend on the background fields. The vectors $\bm{S}_j$ represent the source terms associated to the point-particle stress-energy tensor. The explicit forms of $\bm{V}_{j}$ and $\bm{S}_j$ are derived in Appendix \ref{app:line}, where we also show that the functions $H_2$ and $H_0$ can be completely determined in terms of $\bm{\Psi}_{\rm p}$. For the axial sector, the source terms vanish identically and we can write the coupled first-order equations into a second-order Schr\"odinger-like equation with an effective potential.

\subsection{Computation of the QNMs}\label{sec:qnms}
The late-time gravitational signal from a perturbed BH is dominated by a sum of exponentially damped sinusoids, the QNMs, which correspond to the characteristic vibration modes of the spacetime~\cite{Nollert:1999ji,Berti:2009kk}.
The QNMs are solutions of the sourceless wave equations~\eqref{eq:perturbations}
(i.e., $\bm{S}_j=0$), along with proper boundary conditions---purely outgoing
waves at infinity and ingoing waves at the horizon. 
The latter correspond to the following behavior of the wave function at the
boundaries
\be \bm \Psi_j\propto \op\{ \begin{array}{ll} e^{-i\omega r_*}\,,&r\sim r_h \,,\\
e^{i\omega r_*}\,, &r\to\infty\,, \end{array}\cl.  \ee
where $r_h$ is the horizon radius in these coordinates and the ``tortoise''
coordinate $r_*$ is defined by
\be \frac{dr_*}{dr}=\frac{1}{\sqrt{A B}}\,.  \ee

To compute the QNMs, we use a direct-integration method. We construct a square matrix $\bm X$ (four and two dimensional in the polar and axial case, respectively) of which the columns are independent solutions of Eq.~\eqref{eq:perturbations}. This matrix can be constructed by a certain combination of the solution which is regular at the horizon and the one which is regular at infinity (cf. Refs.~\cite{Pani:2013pma,Macedo:2016wgh} for details). For the polar sector, we note that the boundary conditions defining the QNM eigenvalue problem depend on two parameters, related to the amplitudes of the scalar and gravitational perturbations. 
Two of the columns of $\bm X$ can be constructed by integrating the equations from the horizon outward. Likewise, two other solutions can be constructed by integrating the equations from infinity inward. In general, the solutions integrated from the horizon are linearly independent from the ones integrated from infinity, unless $\omega$ is the QNM frequency. In other words, the QNM frequencies are obtained by imposing 
\be
{\rm det} (\bm X)|_{r=r_{\rm m}}=0\,,
\ee
where $r_{\rm m}$ is an arbitrary matching point of the order of the horizon
radius. The same procedure can be done for the axial modes, with the difference
that the boundary conditions apply only to the gravitational amplitude and therefore the problem is technically
less involved.

We have computed the lowest QNMs of static BHs using both the full numerical background and the perturbative analytical solution given by Eqs.~\eqref{eq:expme}--\eqref{eq:expphi}, up to $N=4$, i.e. up to ${\cal O}(\zeta^6)$. We checked the numerical stability of the QNM frequencies against changes in the values of the numerical horizon $r_h$, numerical infinity and the matching radius $r_{\rm m}$ (typically, we use $r_{\rm m}\sim 4 r_h$).
One of the advantages of the above procedure is that it can be applied to both the perturbative solution and the full numerical one.

\subsection{Plunging particles}\label{sec:plunge}

To obtain the metric and dilaton perturbations due to a particle plunging into a BH, we need to solve the inhomogeneous system~\eqref{eq:perturbations}. Two common methods used in the literature to solve this problem are a direct integration and a Green's function approach (see, e.g., Ref.~\cite{Cardoso:2016olt}). 

The Green's function method relies on the fundamental matrix $\bm X$, as constructed by using the homogeneous solutions as discussed in the previous section. The formal solution of \eqref{eq:perturbations} can be written as \cite{boyce2008elementary}
\be
\bm \Psi_j=\bm X \bm \beta +\bm X\int dr \bm X^{-1}\bm S\,,
\label{eq:formal_sol}
\ee
where $\bm \beta$ is a constant vector to be determined by imposing the proper boundary conditions. Thus, once the fundamental matrix $\bm X$ of the homogeneous problem is computed, its convolution with the source term $\bm S$ in \eqref{eq:formal_sol} yields the solution of the inhomogeneous problem. However, in many problems, the source term might converge slowly at the boundaries or even diverge, as in our case in which the source term diverges at the event horizon. These problems can be avoided by performing a suitable nontrivial transformations of the perturbation functions~\cite{Sasaki:1981kj,Cardoso:2002jr}, or with a careful choice of Green's function~\cite{Poisson:1996ya}.

A different scheme which avoids this problem consists in integrating directly the full inhomogeneous system, by imposing the proper boundary conditions for the full solution. First, we expand the perturbation functions near the horizon as
\be
\bm \Psi_j(r\to r_h)\approx \sum_{k=0}^{N}(r-r_h)^{k+p}\bm \psi_{j,k}e^{-i\omega r_*} +\bm \Psi_{j,H}\,,
\ee
where $p$ is a constant chosen such that the above expansion satisfies the inhomogeneous equations near the horizon and $\bm \Psi_{j,H}$ is the (ingoing) solution of the homogeneous equation. The above expansion is solved iteratively near the event horizon for the coefficients $\bm \psi_{j,k}$ up to $k=N$, and they generically depend on constants (say $\bm \psi_{j,0}$) which are related to the amplitude of the fields at the horizon. The amplitudes at the horizon are then used as shooting parameters; i.e. we chose them such that the numerical solution satisfies the proper boundary conditions also at infinity. For arbitrary amplitudes at the horizon, the numerical solution far from the BH is a combination of ingoing and outgoing waves, i.e.,
\be
\bm \Psi_j=\bm\Psi_j^{out}+\bm\Psi_j^{in}\,,
\ee
and the required solution is obtained by setting the amplitude of the ingoing waves to zero.  In the EDGB case, the amplitudes are related to the gravitational and dilaton perturbations, and therefore the problem is a two-parameter shooting problem for the amplitudes at the horizon. Note that for radial plunging, since the source terms for the axial sector are zero, the particle only induces perturbations in the polar sector.

With the numerical solution at hand, one can compute the gravitational and scalar energy spectra at infinity. This is achieved through the effective stress-energy tensor for the gravitational perturbations (i.e., the Isaacson tensor) and the stress-energy tensor of the dilaton field \cite{Isaacson:1968zza}. As shown in Ref.~\cite{Stein:2010pn}, the Isaacson tensor is the same in GR and in a class of theories including EDGB gravity. The gravitational flux, for a given multipole $l$, can be written as \cite{Zerilli:1971wd}
\be
\frac{dE_g}{d\omega}= \frac{1}{32\pi} \frac{(l+2)!}{(l-2)!}|K(r\to\infty)|^2\,,
\ee
where $K(r)$ is the polar perturbation given by Eq. (\ref{hab13}).~\footnote{Note that, formally, $K$ is a gauge-dependent quantity whereas the use of gauge-independent quantities is obviously desired. Moreover, at large distances the Regge-Wheeler gauge is not well defined.
However, in this limit the function $K$ yields the so-called Zerilli function $Z(r)$~\cite{Zerilli:1971wd} (which is a gauge-invariant quantity~\cite{Moncrief:1974am}) since $K(r\rightarrow\infty)=i\omega Z(r\rightarrow\infty)$.}

The dilaton flux reads
\be
\frac{dE_\phi}{d\omega}=\frac{\omega^2}{16\pi}|\phi_1(r\to\infty)|^2\,,
\ee
where $\phi_1(r)$ is given by the dilaton perturbation in Eq. (\ref{Fourier}).

As shown in Appendix~\ref{app:line}, since the particle does not have a direct coupling to the dilaton field, dilaton radiation only exists due to the gravitational perturbations. In this sense, the gravitational perturbations work as a ``source'' for the dilaton radiation. Since in the dilaton equation they are proportional either to the parameter $\zeta$ or the background scalar field (and derivative), it is natural to expect that the dilaton radiation scales with $\zeta^2$. Additionally, for $l=0$ and $l=1$ the perturbations of EDGB BHs can be written as a single second-order ordinary differential equation, which represents the dilaton perturbation (see Appendix~\ref{app:line}).

\section{Stability and QNMs of EDGB black holes}
%
\begin{figure*}
\includegraphics[width=\columnwidth]{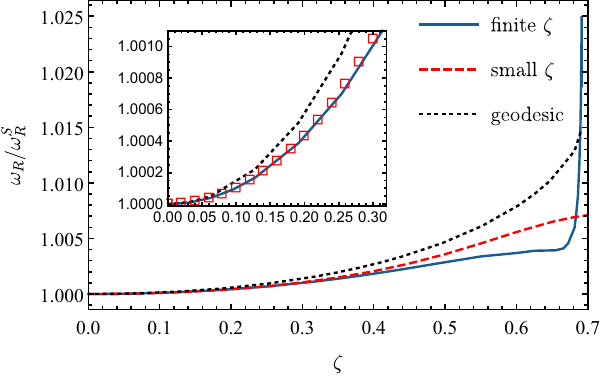}\includegraphics[width=\columnwidth]{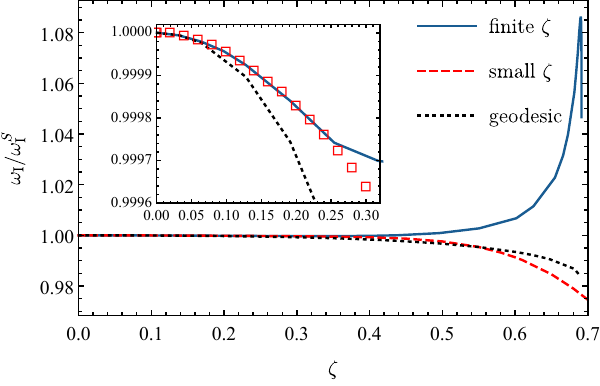}%
\caption{Real (left) and imaginary (right) parts of the axial $l=2$ fundamental mode, normalized by the Schwarzschild ($\zeta=0$) values. We compare three different approaches: the geodesic (black dotted line), the small-coupling limit (red dashed line), and the finite $\zeta$ regime (blue solid line). In the inset we show that for small values of $\zeta$ the small-coupling limit (red squared) agrees well with the finite-coupling result up to $\zeta\approx 0.3$.}%
\label{fig:axial_modes}
\end{figure*}
As previously discussed, the QNMs govern the late-time behavior of any small fluctuation away from axisymmetry of the BH. Since the time dependence is of the form $e^{-i\omega t}$, the {\it absence} of a QNM frequency with the {\it positive imaginary} part in the spectrum implies that all fluctuations decay exponentially with time. Thus, a criterion for linearized mode stability of the spacetime is that all its QNM frequencies have a negative imaginary part~\cite{Berti:2009kk}. In addition, the late-time dynamics is controlled by the fundamental QNM, i.e., the mode with the smallest imaginary component (equivalently, with the longest decay time).

The Schwarzschild spacetime is stable, and its fundamental QNM frequency, $\omega^{S}=M\omega_{R}^{S}+iM\omega_{I}^{S}$, for the $l=2$ mode reads~\cite{Chandrasekhar:1975zza,Berti:2009kk,ref:webpage}
\begin{eqnarray}
 M\omega^{S}&\approx&0.3737-i\,0.08896 \quad {\rm gravitational}\label{eq:qnm_s}\\
 M\omega^{S}&\approx&0.4836-i\,0.09676 \quad {\rm scalar}\label{eq:qnm_ss}
\end{eqnarray}
for the gravitational and scalar fundamental modes, respectively, where $M$ is the BH mass. In EDGB gravity, because the coupling $\zeta$ is smaller than unity, we expect that the fundamental QNM frequencies\footnote{In the $\zeta\to0$ limit the scalar sector is decoupled and we expect to recover the scalar QNMs of a Schwarzschild BH. This is confirmed by the computation presented in this section.} are only slightly different from Eqs.~\eqref{eq:qnm_s} and \eqref{eq:qnm_ss}. In other words, we expect EDGB BHs to be stable for sufficiently small coupling. 
We will now study if these BHs are stable throughout all values of $\zeta$, and also quantify the deviation from the corresponding Schwarzschild value.

We have computed the QNMs of EDGB BHs with two independent codes, both in a small-$\zeta$ expansion and using the full numerical background.
In the small-coupling limit, where the expressions can be expanded in powers of $\zeta$ (see Appendix~\ref{app:motion}), the real and imaginary parts of the QNM frequencies can be written as
\be
\frac{\omega_{R}}{\omega_{R}^{S}}=1+\sum_{j=1}^{N}R_{j}\zeta^j \,,\quad\frac{\omega_I}{\omega_{I}^{S}}=1+\sum_{j=1}^{N}I_{j}\zeta^j \,,
\label{eq:expan_modes}
\ee
where $\omega_{R}^{S}$ and $\omega_{I}^{S}$ are, respectively, the real and imaginary parts of the modes of Schwarzschild BH with the same mass $M$. 
The coefficients $R_j$ and $I_j$ are obtained by fitting the numerical data with the above expression and depend on $l$ and on the nature of the mode.

\subsection{QNMs, light ring, and geodesic correspondence} \label{sec:geodesics}
Computing the QNMs of spacetimes known only numerically can be challenging. For
spherically symmetric spacetimes, a WKB-type analysis has shown that, in the
eikonal ($l\gg 1$) limit, the QNMs can be obtained using only properties of the
light ring (which defines the radius of the photon sphere of the BH). This ``null geodesic
correspondence''~\cite{Ferrari:1984zz,Cardoso:2008bp,Yang:2012he} is useful as
it requires only manipulation of background quantities which are easy to obtain,
and provides a clear physical insight into the QNMs of BHs: they correspond to
waves trapped near the peak of the potential barrier for null particles (i.e., within the photon sphere), slowly
leaking out on a time scale given by the geodesic instability time scale.

The geodesic correspondence only works, formally, in the $l\gg 1$ regime, but can be used even at low $l$ using an appropriate calibration. In fact, this approach has proven to provide reliable results also for low-$l$ modes for a variety of BH spacetimes~\cite{Cardoso:2008bp}, including Kerr-Newman BHs~\cite{Cardoso:2016olt}.

By extending the analysis of Refs.~\cite{Ferrari:1984zz,Cardoso:2008bp,Yang:2012he}, Ref.~\cite{Cardoso:2016olt} recently showed that the complex QNMs of a stationary and axisymmetric BH in the eikonal limit can be written as
\begin{equation}
 \omega_R+i\omega_I \sim \Omega l-i(n+1/2)|\lambda|\,, \label{eikonal}
\end{equation}
where $n$ is the overtone number,
\begin{eqnarray}
 \Omega &=&\frac{-g_{t\varphi}'+\sqrt{{g_{t\varphi}'}^2-g_{tt}' g_{\varphi\varphi}'}}{g_{\varphi\varphi}'}\,,
\end{eqnarray}
is the orbital frequency at the light ring on the orbital plane, and
\begin{eqnarray}
 \lambda = -\frac{1}{\dot t}\sqrt{\frac{V''}{2}}\,, \qquad \dot t=-\frac{E^2 g_{\varphi\varphi}+L g_{t\varphi}}{g_{t\varphi}^2-g_{tt}g_{\varphi\varphi}}
\end{eqnarray}
is the Lyapunov coefficient evaluated at the light-ring location on the equatorial plane. In the above expression, a prime denotes radial derivative, whereas $E$ and $L$ are the (conserved) specific energy and angular momentum of the geodesic, and $V$ is the effective radial potential. The expression~\eqref{eikonal} is valid for $l=m\gg1$ modes. A more involved result for the QNMs of a Kerr BH with generic $l\gg1$ and $|m|\leq l$ is derived in Ref.~\cite{Yang:2012he}.

The geodesic correspondence has been formally proven for Kerr BHs and for a variety of spacetimes, but it has never been checked for BH solutions in modified gravity. This is particularly interesting in light of the breaking of the isospectrality of the axial and polar QNMs of BHs in EDGB theory, as we discuss in the next section. It is important at this stage to point out that the geodesic approximation must fail to capture some of the features of the full problem. It is well adapted, in principle, to describe
the effects of rotation, but cannot take into account (at least not blindly) the presence of extra degrees of freedom like scalars in EDGB. Thus, using this correspondence to extract the QNMs of BHs in modified gravity should be done carefully.

As previously discussed, the background solution and the metric of a spinning BH in EDGB theory is known analytically up to ${\cal O}(\chi^5,\zeta^7)$~\cite{Maselli:2015tta} and numerically for any value of $\chi$ and $\zeta$~\cite{Kleihaus:2011tg,Kleihaus:2015aje}, where $\chi$ is the dimensionless angular momentum parameter,
\be
\chi=J/M^2\,.
\ee
We will use this result in Sec. \ref{detectability}, to estimate the modes of spinning EDGB BHs.

\subsection{Axial modes}

\begin{table*}%
\caption{Numerical value of the coefficients $R_j$ and $I_j$ for the expansions in the small-coupling limit, cf.~\eqref{eq:expan_modes} for the axial QNMs. The geodesic coefficients are computed from the exact analytical solution for small $\zeta$ limit, while the QNM frequencies coefficients are obtained through a polynomial fit with the data.}
\label{tab:coeff_axial}
\begin{tabular}{ccccc}
\hline\hline
  & $j$  & $l=2$  & $l=3$ & Geodesic \\
\hline
\multirow{4}{*}{$R_j$} 
 & 1    &  0 					& 0 & 0 \\
 & 2    &  $1.002\times10^{-3}$  & $1.173\times 10^{-2}$  & $1.257\times 10^{-2}$  \\
 & 3    &  $1.906\times 10^{-3}$ & $5.035\times 10^{-3}$  & $6.872\times 10^{-3}$  \\
 & 4    &  $1.131\times 10^{-3}$ & $1.353\times 10^{-2}$  & $5.537\times 10^{-3}$ \\ \hline
\multirow{4}{*}{$I_j$} 
 & 1    &	 0  & 0 & 0\\
 & 2    & $-5.174\times 10^{-3}$ & $-4.774\times 10^{-3}$  & $-5.267\times 10^{-3}$ \\
 & 3    & $5.766\times 10^{-3}$  & $7.590\times 10^{-4}$   & $-7.184\times 10^{-3}$ \\
 & 4    & $-7.091\times 10^{-3}$ & $-3.282\times 10^{-3}$  & $-7.822\times 10^{-3}$ \\ 
\hline \hline
\end{tabular}
\end{table*}

The axial sector of gravitational perturbations is decoupled from the scalar-field perturbations, and hence is simpler to study. Using the direct-integration procedure described in Sec.~\ref{sec:qnms}, we have computed the axial QNMs both in a small-$\zeta$ expansion and in the full numerical background. Our results are summarized in Fig.~\ref{fig:axial_modes} and in Table~\ref{tab:coeff_axial}. 

In Fig.~\ref{fig:axial_modes}, we show the behavior of the axial $l=2$
fundamental mode as a function of the coupling $\zeta$, normalized by the
corresponding Schwarzschild quantity.  In most of the range of $\zeta$, the
behavior of the modes is smooth and given by a corresponding small deformation
of the Schwarzschild QNMs. The only exception occurs close to the critical value
of the coupling constant $\zeta \approx 0.691$, where the QNMs have a very
sensitive dependence on $\zeta$. For small values of $\zeta$, say
 for $\zeta\lesssim 0.4$,  analytically
expanded backgrounds [up to ${\cal O}(\zeta^6)$] yield QNMs which are in very
good agreement with the full numerical solution. This provides a nontrivial
check for both our (independent) codes.

Figure~\ref{fig:axial_modes} also shows the result of the geodesic algorithm
described in Sec.~\ref{sec:geodesics}.
For the axial modes, which are decoupled from the scalar perturbations,
the geodesic predictions are in good agreement with the results of 
the full numerical solution.
Although not shown in Fig.~\ref{fig:axial_modes}, the agreement is better for higher multipoles, as
expected (cf. Table~\ref{tab:coeff_axial}).

Finally, Table~\ref{tab:coeff_axial} shows the results of the polynomial
fit~\eqref{eq:expan_modes} to the axial QNM of EDGB BHs, which is specially
accurate for small $\zeta$. By analyzing the perturbation equations, it is easy
to show that $R_1=I_1=0$ in the expansion~\eqref{eq:expan_modes}. The full
numerical results are available online~\cite{ref:webpage}.
In Table~\ref{tab:coeff_axial}, we also give the coefficients obtained by the geodesic algorithm, which help to quantify the accuracy of the geodesic approximation; for instance at $\zeta=0.5$ the difference in percentage between the real and the 
imaginary parts of the (not-normalized) frequency, relative to the numerical result,
is, respectively, less than $3\%$ and less than $8\%$ for $l=2$, and improves for $l>2$. Similar deviations are obtained in the GR limit, $\zeta=0$. 

As explained in Appendix~\ref{app:line}, similarly to what happens in GR, there are no axial QNMs for $l=0$ and $l=1$.

\begin{figure*}[ht]%
\includegraphics[width=\columnwidth]{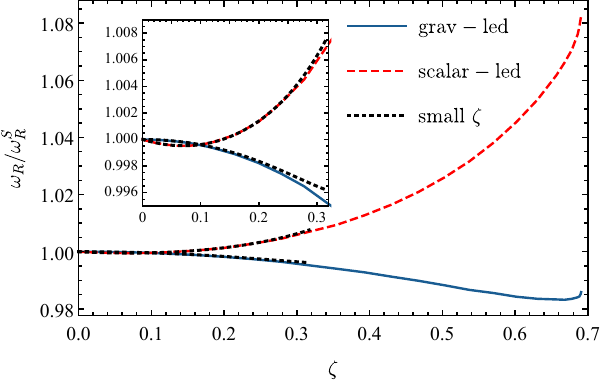}\includegraphics[width=\columnwidth]{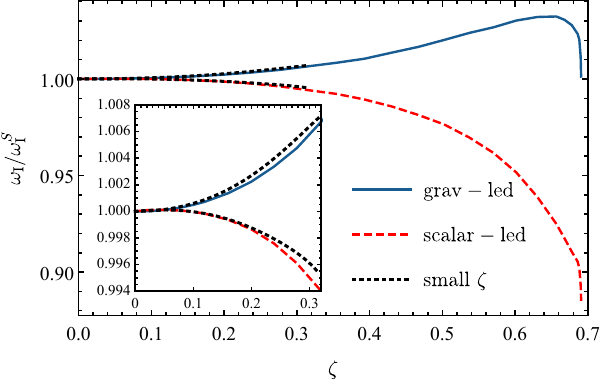}%
\caption{Real (left) and imaginary (right) parts of the polar quasinormal modes for $l=2$, for the gravitational- and scalar-led modes, as functions of the coupling $\zeta$, normalized by the Schwarzschild-limit quantities. The insets show a closeup in order to see the comparison for small values of $\zeta$.
}%
\label{fig:polar_modes}%
\end{figure*}
%
\subsection{Polar modes}

%
\begin{table}%
\caption{Numerical value of the coefficients $R_j$ and $I_j$ for the polar gravitational-led and scalar-led modes. 
}
\label{tab:coeff}
\begin{tabular}{c c c c}
\hline\hline
  & $j$ & {\footnotesize Polar, gravitational $l=2$} & {\footnotesize Polar, scalar $l=2$}  \\
\hline
\multirow{4}{*}{$R_j$} 
 & 1 & 0                     & $-1.408\times 10^{-2}$ \\
 & 2 & $-3.135 \times 10^{-2}$ & $1.127\times 10^{-1}$  \\
 & 3& $-9.674\times 10^{-2}$ & $-1.462\times 10^{-1}$  \\
 & 4 & $2.375\times 10^{-1}$ & $5.334\times 10^{-1}$\\ \hline
\multirow{4}{*}{$I_j$} 
 & 1 & 0                  &$5.580\times 10^{-3}$\\
 & 2  & $4.371\times 10^{-2}$ &$-6.780\times 10^{-2}$ \\
 & 3  & $1.794\times 10^{-1}$ &$1.042\times 10^{-1}$\\
 & 4  & $-2.947\times 10^{-1}$ & $-2.868\times 10^{-1}$\\ 
\end{tabular}
\begin{tabular}{c c c c}
\hline\hline
  & $j$ & {\footnotesize Polar, gravitational $l=3$}  & {\footnotesize Polar, scalar $l=3$}   \\
\hline
\multirow{4}{*}{$R_j$} 
 & 1 & 0                     & $-6.361 \times 10^{-3}$ \\
 & 2 & $-9.911\times 10^{-2}$ & $1.442\times 10^{-1}$  \\
 & 3& $-4.907 \times 10^{-2}$ & $1.168\times 10^{-1}$  \\
 & 4 & $9.286\times 10^{-2}$ & $-1.803\times 10^{-1}$\\ \hline
\multirow{4}{*}{$I_j$} 
 & 1 &  0                  &$2.906\times 10^{-3}$\\
 & 2  & $7.710\times 10^{-2}$ &$-5.670\times 10^{-2}$ \\
 & 3  & $1.399\times 10^{-1}$ &$-1.445\times 10^{-1}$\\
 & 4  & $-3.450\times 10^{-1}$ & $2.105\times 10^{-1}$\\ 
\hline \hline
\end{tabular}
\end{table}

Unlike the axial sector, the polar gravitational sector of the metric perturbations couples to the scalar-field perturbations.
The system of ODEs is more complex and finding the QNM frequencies is therefore more challenging. Even for arbitrarily small values of $\zeta$ the QNMs contain two families: (i) \emph{gravitational-led} modes, which reduce to the gravitational QNMs of Schwarzschild BHs in the $\zeta\to0$ limit, and (ii) \emph{scalar-led} modes, which reduce to the QNMs of a test scalar field on a Schwarzschild metric when $\zeta\to0$ (see  Ref.~\cite{Molina:2010fb} for a similar situation in another theory). The extent to which each of these modes is excited in actual physical setups
is discussed in the next section.

Our results for the quadrupole modes ($l=2$) are summarized in Fig.~\ref{fig:polar_modes} and
Table~\ref{tab:coeff}, where we show the fundamental gravitational-led and
scalar-led QNMs. The deviations from the GR case are larger than in the axial
case. This is probably due to the extra coupling between gravitational and
scalar degrees of freedom.  From the fits given in Table~\ref{tab:coeff}, it is
interesting to note that the leading-order correction to the $l=2$
gravitational-led mode has an opposite sign compared to the axial mode. In
particular, since the geodesic correspondence predicts only one type of modes
and the latter are in good agreement with the axial modes, we find that the
behavior of the polar modes is not captured by the geodesic correspondence; therefore, we do not plot the geodesics results in the figure.
This qualitative difference is
expected, because the axial potential resembles the geodesic potential at large
$l$, whereas in the polar sector the coupling to scalar perturbations
drastically changes the dynamics of the perturbations. Likewise, there is no
reason to expect that the behavior of scalar-led perturbations is well captured
by the geodesic correspondence, at least for small values of $l$. 

Due to the coupling between the dilaton and gravitational perturbations, there are also nontrivial $l=0,\,1$ scalar-led modes for EDGB BHs. These reduce to their respective scalar modes in the Schwarzschild spacetime when $\zeta\to 0$. The results at finite coupling follow the trend of higher multipoles.

\subsection{Mode stability}

%
From the above results, it is clear that the fundamental QNMs of an EDGB BH change at most by a few percent relative to the Schwarzschild case. As a consequence, these modes are stable for any value of $\zeta$ in the domain of existence of static EDGB BHs. We have investigated this issue also for higher multipoles ($l\geq2$) and our numerical search has found no unstable modes in the entire parameter space. This strongly indicates that static EDGB BHs are linearly mode stable, just like Schwarzschild BHs.

\section{Radial plunge}
In this section, by using the procedures depicted in Sec.~\ref{sec:plunge}, we discuss the gravitational and dilaton radiation emitted by a particle plunging radially into an EDGB BH. For practical reasons, this computation was done within the small-$\zeta$ approach for the background, and the equations are expanded up to order ${\cal O} (\zeta^6)$. As discussed in the previous section, this higher-order perturbative approach gives precise results even for relatively large values of $\zeta$.

\begin{figure}%
\includegraphics[width=\columnwidth]{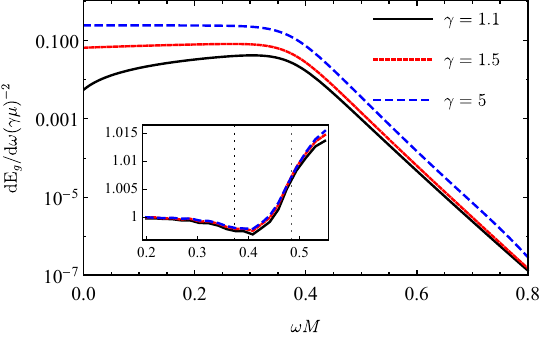}%
\caption{Gravitational quadrupolar flux for radial plunges into an EDGB BH with different boosts for $\zeta=0.1$. For small couplings the changes in the gravitational fluxes are very small. In the inset, we plot the ratio of the fluxes for $\zeta=0.1$ and for the Schwarzschild spacetime. The vertical dotted lines in the inset are the values of the gravitational- and scalar-led QNMs.}
\label{fig:flux_g}%
\end{figure}
\begin{figure}%
\includegraphics[width=\columnwidth]{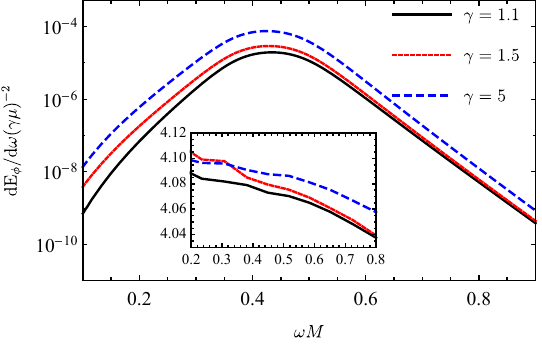}
\caption{Dilaton quadrupolar flux for radial plunges into an EDGB BH with different boosts for $\zeta=0.1$. As expected, the scalar flux starts to present an exponential suppression for frequencies roughly larger than the dilaton mode. In the inset, we plot the ratio between the fluxes for $\zeta=0.1$ and $\zeta=0.05$, which confirms that the scalar flux scales dominantly with $\zeta^2$. 
}%
\label{fig:flux_phi}%
\end{figure}

In Fig.~\ref{fig:flux_g} we show the gravitational flux, by considering different
initial boosts for the particle (see Appendix~\ref{app:motion}). 
The deviations from the GR case are very small,
of order of $\sim 1\%$, at least for $\zeta\lesssim0.1$. 
Moreover, we note that all additional terms appearing in the metric perturbation
equations (see Appendix~\ref{app:line})---sources included---are at least of
order $\sim\zeta^2$, and therefore the corrections to the flux are proportional
to $\sim\zeta^2$. Therefore, in the small-coupling limit, the gravitational flux
can be written as
\be \frac{dE_g}{d\omega}\sim \frac{dE_g^{S}}{d\omega}\op[1+{\cal
O}(\zeta^2)\cl]\,, \ee
where $dE_g^{S}/d\omega$ is the corresponding Schwarzschild flux.

Although the overall gravitational corrections are small, the dilaton perturbations can also be radiated by the plunging particles, similarly to the case of a neutral particle plunging into a charged BH~\cite{Johnston:1973cd,Johnston:1974vf,Cardoso:2016olt}. This is due to the coupling between the gravitational and dilaton perturbations, cf. Eq.~\eqref{eq:dilaton_per}. We show the dilaton flux for $\zeta=0.1$ in Fig.~\ref{fig:flux_phi}. Note that the flux displays a cutoff roughly at $\omega\sim\omega_R^{\phi}$, where $\omega_R^{\phi}$ is the scalar-led QNM. Also in this case the dilaton flux scales dominantly as $\zeta^2$. Additionally, because the source terms in the dilaton field are only due to the gravitational perturbations, the dilaton radiation is also dominantly quadrupolar (see Ref.~\cite{Johnston:1973cd} for a similar setup with perturbations in Reissner-Nordstr\"om BHs).

Although the total radiated energy due to the dilaton radiation scales as
$\zeta^2$, it can still be considerably high, depending on the radiating source. For
instance, for the source GW150914 the luminosity due to the GWs was
${dE_g}/{dt}\approx 3.6 \times 10^{56}{\rm erg/s}$~\cite{Abbott:2016blz}.
Therefore, even if the scalar radiation is small compared to the gravitational
one, it can still have a considerable value. The implication of a burst of
dilaton radiation depends on how the environment (plasma, surrounding stars,
 etc.) interacts with the dilaton field.
%

%

\section{Constraints on the EDGB coupling from ringdown observations\label{detectability}} 
The results of the previous section show that plunges of point particles do not excite considerably the scalar-led QNMs. This suggests that such modes might be only mildly excited during the coalescence of two BHs of equal mass and that the main signature of EDGB theory would be a shift of the ringdown frequencies, the latter being governed by the fundamental gravitational-led modes.
Thus, the use of ringdown measurements in EDGB theory to estimate the magnitude of the coupling 
$\zeta$ would rely only on the deviation of the gravitational-led modes from their GR counterpart. These deviations are parametrized in terms of the fit~\eqref{eq:expan_modes} of which the coefficients are given in Tables~\ref{tab:coeff_axial} and \ref{tab:coeff} for the $l=2,3$ fundamental axial and polar modes, respectively.

The results of the previous sections refer to nonspinning BHs, whereas the end product of the coalescence is a spinning compact object. Thus, ringdown tests require the knowledge of the first dominant modes of a spinning BH as a function of the spin $\chi$ and of the coupling constant of the modified theory of gravity. Computing the QNMs of generic spinning BHs in modified gravity is a very challenging task\footnote{The task of computing the QNMs of a Kerr BH is enormously simplified by the fact that the gravitational perturbation equations are separable, due to
special properties of the background Kerr geometry which does not necessarily hold for other spinning BH solutions.}, which has witnessed some developments only recently (cf. Ref.~\cite{Pani:2013pma} for an overview). Nonetheless, most of the results are obtained within a perturbative expansion valid for $\chi\ll1$ and quickly become intractable at the higher perturbative order. The latter is required to extrapolate the perturbative result up to $\chi\approx 0.7$, which is roughly the spin of the final BH measured in the two coalescence events  detected by aLIGO to date~\cite{Abbott:2016blz,Abbott:2016nmj}.

To overcome this limitation and estimate how rotation affects the QNMs of an EDGB BH, we rely on the geodesic correspondence and on our knowledge of the metric of a spinning BH in EDGB theory. As previously mentioned, the latter is known analytically up to ${\cal O}(\chi^5,\zeta^7)$~\cite{Maselli:2015tta} and numerically for any value of $\chi$ and $\zeta$~\cite{Kleihaus:2011tg,Kleihaus:2015aje}.
In the previous section we have checked that the geodesic correspondence works reasonably well for axial modes, so in this section we will focus on the latter. However, we may argue that the order of magnitude of our estimates should be correct also for the more relevant polar modes. 

By using the geodesic correspondence for $l=m$ modes described in Sec.~\ref{sec:geodesics}, we obtain the following result 
\begin{widetext}
 \begin{eqnarray}
 \frac{\omega_R}{\omega_R(\chi=0)} &=& 1\, +\left(0.3849\, +0.0326 \zeta ^2\right) \chi +\left(0.2038\,
   +0.0264 \zeta ^2\right) \chi ^2+\left(0.1283\, +0.0169 \zeta^2\right) \chi ^3\nn\\
   &&+\left(0.0897\, +0.0105 \zeta ^2\right) \chi^4+\left(0.0671\, +0.0054 \zeta ^2\right) \chi ^5+{\cal O}(\chi^6,\zeta^3)\,,\\
 \frac{\omega_I}{\omega_I(\chi=0)} &=& 1\, -0.0059 \zeta ^2 \chi -\left(0.0741+0.0066 \zeta ^2\right)
   \chi ^2-\left(0.0713-0.0002 \zeta ^2\right) \chi^3\nn\\
  &&-\left(0.0604-0.0050 \zeta ^2\right) \chi^4-\left(0.0504-0.0079 \zeta ^2\right) \chi^5 +{\cal O}(\chi^6,\zeta^3)\,,\label{geodesic_correspondence}
\end{eqnarray}
\end{widetext}
where $\omega_{R,I}(\chi=0)$ are the corresponding axial modes for a nonspinning EDGB BH shown in Table~\ref{tab:coeff_axial}.
The above expressions are valid up to ${\cal O}(\chi^5,\zeta^2)$ and for any $l=m\gg1$; the only difference enters in the normalization factors on the left-hand side.

The detection of two ringdown modes is necessary to estimate the mass, spin and
coupling $\zeta$.  Here we adopt the same Fisher-matrix technique presented in
Ref.~\cite{Cardoso:2016olt}. Let us consider first the axial modes, which are, generically,
excited during the merger phase.
The measurement of the two most dominant modes (which we take to be $l=m=2,3$, excited with a relative amplitude of $3:1$~\cite{Berti:2007zu}) gives us sufficient information to extract the mass and spin of the BH, as well as the coupling parameter of the theory. To do this, we use the geodesic correspondence, Eq.~\eqref{geodesic_correspondence}. {\it Assuming} that both modes are excited to detectable amplitude (this can be quantified using the methods of Ref.~\cite{Berti:2007zu}), then a Fisher-matrix computation allows us to estimate the uncertainties in the parameters determining the ringdown~\cite{Cardoso:2016olt}: $M, \chi, \zeta$. This then allows us to constrain the magnitude of the coupling parameter $\zeta$,
\begin{equation} \zeta \lesssim \frac{3.98 -0.718 \chi+0.181\chi^2-0.045
\chi^3}{\sqrt{\rho}}\,, \label{boundzeta} \end{equation}
where $\rho$ is the signal-to-noise ratio in
the ringdown waveform and the ${\cal O}(\chi^4)$ and ${\cal O}(\chi^5)$ terms
are negligible. The numerator in the expression above is a (mildly) decreasing
function of the spin and ranges from $\approx 4$ to $\approx 3.5$ in the region
$0\leq\chi\approx0.8$.

This analysis can be extended to polar modes. In fact, one might even argue that corrections to polar modes are higher, since the polar QNMs of nonrotating BHs are more affected than the axial, cf. Tables~\ref{tab:coeff_axial} and \ref{tab:coeff}.
However, there is little evidence that they follow the geodesic correspondence, 
but we expect that the order of magnitude change in the modes remains the same.
Therefore, it is reasonable to believe that, even when polar modes are excited, a mode analysis of the ringing signal can set constraints on $\zeta$ of the order of those given by Eq.~(\ref{boundzeta}).

As a reference, the final BH spin of GW150914 was $\chi\approx0.67$~\cite{Abbott:2016blz} and $\rho\approx7.7$ in the ringdown part~\cite{TheLIGOScientific:2016src}.
This [assuming, as a first approximation, that Eq.~(\ref{boundzeta}) also holds for polar modes] yields roughly $\zeta\lesssim 1.3$, which is weaker than the theoretical bound~\cite{Pani:2009wy} for the finite-$\zeta$ solution, $\zeta\lesssim0.691$, and therefore meaningless.
This constraint is nevertheless comparable with that derived from the orbital decay rate of low-mass x-ray binaries~\cite{Yagi:2012gp}
and it is slightly larger than the projected bound achievable in the near future from the measurements of quasiperiodic oscillations in the spectrum of accreting BHs~\cite{Maselli:2014fca}, although the latter might be affected by astrophysical systematics.
Our estimate suggests that $\rho\gtrsim25$ in the ringdown waveform is needed to obtain an upper limit which is more stringent than the theoretical bound using GW ringdown detections.
Of course, in order to set these upper limits, an accurate computation of both polar and axial modes of rotating EDGB BHs will be needed.
\section{Discussion and conclusions}
EDGB gravity is a simple and viable higher-curvature correction to GR which predicts BH solutions with scalar charge. Since strong-curvature corrections are suppressed at large distance, it is natural to expect that the most stringent constraints on this theory come from the strong-curvature, highly dynamical regime as the one involved in a BH coalescence. Furthermore, compact stars in this theory possess only a very small scalar charge~\cite{Yagi:2011xp,Kleihaus:2016dui} and therefore EDGB gravity evades the stringent constraints on the dipole radiation coming from current binary-pulsar systems~\cite{Berti:2015itd}.

The estimate~\eqref{boundzeta} translates to the following upper bound on the dimensionful EDGB coupling\footnote{Since one of the parameters of our Fisher-matrix analysis is $\zeta=\alpha/M^2$, propagation of errors implies a relative uncertainty $\delta\alpha/\alpha=\delta\zeta/\zeta+2\delta M/M$. However, in the large-$\rho$ limit the term $\delta M/M$ is negligible because it scales as $1/\rho$, compared to the $1/\sqrt{\rho}$ behavior of the error on $\zeta$ [cf. Eq.~\eqref{boundzeta}]. Therefore, in this limit $\delta \alpha\lesssim \delta \zeta M^2$.},
\begin{equation}
 \alpha^{1/2}\lesssim	11 \left(\frac{50}{\rho}\right)^{1/4}\left(\frac{M}{10 M_\odot}\right)\,{\rm km}\,, \label{boundalpha}
\end{equation}
where the prefactor changes by less than $10\%$ depending on the final BH spin. This result is in agreement with the simple estimates derived in Ref.~\cite{Barausse:2014tra}. As a consequence, our analysis also confirms that in most cases modified-gravity effects can be distinguished from environmental effects~\cite{Barausse:2014tra}.

Future GW detectors will greatly increase the signal-to-noise ratio, a large value of which is necessary to perform ringdown tests of the Kerr metric~\cite{Berti:2016lat}. The signal-to-noise ratio of a ringdown waveform scales approximately (among its dependence on other quantities not shown here) as $\rho\sim M^{3/2}/S_n(f)^{1/2}$~\cite{Berti:2005ys}, where $M$ is the final BH mass and $S_n(f)$ is the detector noise power spectral density at a given frequency $f$. The best sensitivity of the future Voyager~\cite{Voyager} and Einstein Telescope~\cite{ET} detectors will be, respectively, roughly a factor of $10$ and a factor of $100$ better than in the first aLIGO observing run at the same optimal frequency $f\sim 10^2\,{\rm Hz}$~\cite{Berti:2016lat}.
Thus, the Einstein Telescope with an optimal design can achieve a signal-to-noise ratio of roughly $\rho\approx 100$ for the ringdown signal of a GW150914-like event. 
From Eqs.~\eqref{boundalpha} and~\eqref{boundzeta}, this would translate into the bound $\alpha^{1/2}\lesssim 8 \left(\frac{M}{10 M_\odot}\right)\,{\rm km}$ and $\zeta\lesssim 0.4$. As expected, lighter BHs would provide a significantly more stringent constraint on $\alpha$, although their ringdown frequency might not fall into the optimal frequency range for ground-based detectors. Due to the small exponent of $\rho$ in Eq.~\eqref{boundalpha}, even an increase of $\rho$ of 1 order of magnitude will not provide a significantly more stringent constraint on the EDGB coupling. A stronger constraint may be set if future observations detect a light BH with a very large signal-to-noise ratio.

Given this scenario, electromagnetic observations of accreting BHs (like the one discussed in Ref.~\cite{Maselli:2014fca}) might provide more stringent constraints in the future, although the latter are affected by astrophysical systematics that are absent in the ringdown case.

Our estimates in the case of spinning BHs rely on the geodesic analogy for QNMs, which we verified only for axial modes in the static case and for Kerr BHs with any spin~\cite{Cardoso:2016olt}. It would be interesting to compute the modes of slowly rotating EDGB BHs (e.g. by adapting the methods discussed in Ref.~\cite{Pani:2013pma}) and to check the geodesic approximation in the spinning case. This computation will be required to place precise constraints on the EDGB coupling through future detections of BH ringing with high signal-to-noise ratio.

Another interesting extension of our work concerns the scalar waves emitted during the coalescence. Although the luminosity in scalar waves is significant, this radiation may be possibly detected only if the dilaton is coupled to matter. Such coupling is presumably small and would not give rise to any effects in the detectors. Nonetheless, if the dilaton-matter coupling is non-negligible, the scalar radiation might be investigated through the same techniques developed to study the scalar emission in scalar-tensor theories, e.g. by using a network of ground-based detectors~\cite{hayama}.

\begin{acknowledgments}
We thank Hector Silva for pointing out misprints in the numerical Table~\ref{tab:coeff}.
V.C. acknowledges financial support provided under the European Union's H2020 ERC Consolidator Grant ``Matter and strong-field gravity: New frontiers in Einstein's theory'' Grant No. MaGRaTh--646597.
C.M. acknowledges financial support from Conselho Nacional de Desenvolvimento Cient\'ifico e Tecnol\'ogico through Grant No.~232804/2014-1.
J.L.B.S., F.S.K. and J.K. gratefully acknowledge support by the DFG Research Training Group 1620 ``Models of Gravity''.
J.L.B.S. and J.K. gratefully acknowledge support by the grant FP7, Marie Curie Actions, People, International Research Staff Exchange Scheme (IRSES-606096).
Research at Perimeter Institute is supported by the Government of Canada through Industry Canada and by the Province of Ontario through the Ministry of Economic Development and Innovation.
This project has received funding from the European Union's Horizon 2020 research and innovation program under the Marie Sklodowska-Curie Grant No.~690904 and from FCT-Portugal through the Project No. IF/00293/2013.
This work was supported by the H2020-MSCA-RISE-2015 Grant No. StronGrHEP-690904.

\end{acknowledgments}

\appendix

\section{Spherically symmetric BHs in EDGB}
\label{app:motion}
\subsection{Spherically symmetric BHs}

{The ansatz for the static EDGB BH is given in terms of the functions $A(r)$ and $B(r)$ for the line element (\ref{eq:linel}), in addition to the function $\phi_0(r)$ for the dilaton field.} 

{At spatial infinity $r=\infty$, the dilaton vanishes and we recover the metric of a flat spacetime. Asymptotically, the functions present the following behavior:}

%
\begin{align}
	&A\sim1-\frac{2 M}{r}+\mathcal{O}(r^{-3}),\\
	&B\sim1-\frac{2M}{r}+\frac{Q^2}{4 r^2}+\mathcal{O}(r^{-3}),\\
	&\phi_0\sim\phi_\infty+\frac{Q}{r}+\frac{M Q}{r^2}+\mathcal{O}(r^{-3}),
\end{align}
where $M$ is the ADM mass of the BH and $Q$ is the ``charge'' of the scalar field\footnote{Note, however, that this is not an independent parameter, and therefore should be considered as a secondary hair~\cite{Kanti:1995vq}.}. At the BH horizon $r=r_h$, we find that the metric functions and the scalar field behave as
\begin{align}
	&A\sim \sum_j^{\infty} a_j(r-r_h)^{j+1},~
	B\sim \sum_j^{\infty}  b_j(r-r_h)^{j+1},\\
	&\phi_0\sim \sum_j^{\infty}  \phi_j(r-r_h)^{j}.
\end{align}
{The coefficients $a_j$, $b_j$ and $\phi_j$ are constants, but they are not free parameters. In fact they depend only on the coupling parameters of the theory and horizon radius of the BH (related to the total mass $M$) through some complicated algebraic relations~\cite{Kanti:1995vq}.} 

{In order to study the quasinormal modes of the full geometry, we build numerically the solutions of the background metric. We do so by integrating the system of ordinary differential equations for $A(r)$, $B(r)$ and $\phi_0(r)$, from the horizon up to infinity. We perform this integration in a compactified coordinate $x = 1-r_h/r$, with $x \in [0,1]$. With a suitable reparametrization of the functions we can impose as boundary conditions the correct behavior at infinity ($x=1$) and at the horizon ($x=0$). The integration is performed with the package COLSYS \cite{Ascher:1979iha}, and typically background solutions are generated with 1000--10000 points and required relative precisions of the functions smaller than $10^{-6}$.}

Another approach, which allows us to obtain analytical expressions for the background metric and dilaton field is the small-coupling limit. In the small-coupling limit, in which $\zeta\equiv\alpha/M^2\ll 1$, the equations can be greatly simplified, and the background field can be computed analytically. Considering the expansion for the background fields
\begin{align}
A&=1-\frac{r_h}{r}+\sum_{j=0}^{N}\zeta^{j+2}A^{\rm c}_{j},\label{eq:expme}\\
B&=1-\frac{r_h}{r}+\sum_{j=0}^{N}\zeta^{j+2}B^{\rm c}_{j},\\
\phi_0&=\sum_{j=0}^{N}\zeta^{j+1}\phi^{\rm c}_{j}\label{eq:expphi}
\end{align}
and expanding the background equations in $\zeta$, one has that each order $j+2$ for the metric and $j+1$ for the scalar field can be solved consistently, imposing regularity at the horizon $r=r_h$ and that the scalar field goes to zero at infinity. Typically, we consider solutions up to $N=4$, verifying that the results converge in the small-$\zeta$ region.

\section{Perturbations}
\label{app:line}
\subsection{Polar sector}
For the polar sector, the nonvanishing components of the modified Einstein's equations are
\begin{widetext}
\begin{align}
&K'+\frac{\alpha  (B-1) e^{\phi _0} }{r^2 \left(\alpha  B e^{\phi _0} \phi _0'-r\right)}\phi _1'+ \left(\frac{1}{r}-\frac{A'}{2 A}\right)K
-i\frac{ (\Lambda +1)}{r^2 \omega }H_1+\frac{ \left(\alpha  (3 B-1) e^{\phi _0} \phi _0'-2 r\right)}{2 r \left(r-\alpha  B e^{\phi _0} \phi _0'\right)}H_2\nn\\
&+\frac{ \left(\alpha  (B-1) e^{\phi _0} \left(r A'+A \left(2-2 r \phi _0'\right)\right)+A r^3 \phi _0'\right)}{2 A r^3 \left(r-\alpha  B e^{\phi _0} \phi _0'\right)}\phi _1
=\frac{4 \sqrt{2} \pi  r A_{{lm}}^{{(1)}}}{r \omega -\alpha  B \omega  e^{\phi _0} \phi _0'},\label{eq:ein_eq1}\\
&H_1'+{\left(\frac{2 (B-1) B \left(r A'-2 A\right)}{A r}-\frac{r \left(B^2 r\phi _0'^2+2 (B-1) B'\right)}{r-\alpha  B e^{\phi _0} \phi _0'}-4 B'\right)}\frac{H_1 }{4 (B-1) B}+\frac{i \omega }{B}H_2 \nn\\
&-{i r \omega  \left(\frac{4 B'}{B}+\frac{r^2 \phi _0'^2}{r-\alpha  B e^{\phi _0} \phi _0'}\right)}\frac{K }{4 (B-1)}-\frac{i \alpha  \omega  e^{\phi _0} B'}{B r^2-\alpha  B^2 r e^{\phi _0} \phi _0'} \phi _1=0,\label{eq:ein_eq2}\\
&H_0'+\frac{r\left(\alpha  B e^{\phi _0} A' \phi _0'-2 A\right)}{2 A \left(r-\alpha  B e^{\phi _0} \phi _0'\right)} K'+{\left(\frac{2 \left(A-r A'\right)}{r-\alpha  B e^{\phi _0} \phi _0'}+3 A'\right)}\frac{H_2 }{2 A} +\frac{\alpha  B e^{\phi _0} A' \left(r \phi _0'-2\right)-A r^2 \phi _0'}{A r^2 \left(r-\alpha  B e^{\phi _0} \phi _0'\right)}\phi _1 \nn\\
&+\frac{\alpha  B e^{\phi _0} A' }{A r^2-\alpha  A B r e^{\phi _0} \phi _0'}\phi _1'+\left(\frac{A'}{2 A}-\frac{1}{r}\right)H_0 +\frac{i  \omega }{A}H_1=0,\label{eq:ein_eq3}\\
&H_2 \left(2 A-3 \alpha  B^2 e^{\phi _0} A' \phi _0'\right)-B r K' \left(A' \left(r-3 \alpha  B e^{\phi _0} \phi _0'\right)+2 A\right)+K \left(\alpha  B e^{\phi _0} \phi _0' \left(2 r \omega ^2-\Lambda  A'\right)+2 A \Lambda -2 r^2 \omega ^2\right)\nn\\
&+\frac{\phi _1 \left(\alpha  e^{\phi _0} \left(B \left(2 r \omega ^2-A' \left((1-3 B) r \phi _0'+3 B+2 \Lambda +1\right)\right)-2 r \omega ^2\right)-A B r^2 \phi _0'\right)}{r^2}\nn\\
&+\phi _1' \left(\frac{\alpha  B (3 B-1) e^{\phi _0} A'}{r}+A B r \phi _0'\right)-\frac{2 A H_0 (\Lambda +1) \left(r-\alpha  B e^{\phi _0} \phi _0'\right)}{r}+A B H_0' \left(2 r-\alpha  (3 B-1) e^{\phi _0} \phi _0'\right)\nn\\
&+2 i B H_1 \omega  \left(2 r-\alpha  (3 B-1) e^{\phi _0} \phi _0'\right)=-16 \pi  A B r^2 A_{{lm}},\label{eq:ein_eq4}\\
&H_2-\frac{A H_0 \left(4 B' \left(r-\alpha  B e^{\phi _0} \phi _0'\right)+B r^2 \phi _0'^2\right)}{2 (B-1) \left(\alpha  B e^{\phi _0} A' \phi _0'-2 A\right)}+\frac{\alpha  e^{\phi _0} \phi _1 \left(A A' B'-B \left(\left(A'\right)^2-2 A A''\right)\right)}{A r \left(\alpha  B e^{\phi _0} A' \phi _0'-2 A\right)}=0,\label{eq:ein_eq5}\\
&\frac{H_2 \left(4 \alpha  B e^{\phi _0} \phi _0' \left((B-1) (\Lambda +1)-r B'\right)+4 r \left(B \left(r B'+B-\Lambda -3\right)+\Lambda +2\right)+B^2 r^3 \phi _0'^2\right)}{4 (B-1) B r^2 \left(r-\alpha  B e^{\phi _0} \phi _0'\right)}\nn\\
&+\frac{K' \left(\frac{r \left(B^2 (-r) \phi _0'^2-2 (B-1) B'\right)}{r-\alpha  B e^{\phi _0} \phi _0'}-4 B'+\frac{8 (B-1) B}{r}\right)}{4 (B-1) B}+\frac{\Lambda  K \left(\frac{4 B'}{B}+\frac{r^2 \phi _0'^2}{r-\alpha  B e^{\phi _0} \phi _0'}\right)}{4 (B-1) r}\nn\\
&+\frac{\phi _1' \left(B r^3 \phi _0'-\alpha  e^{\phi _0} \left(B \left(3 r B'+4 (B-1) r \phi _0'-4 B+4\right)-r B'\right)\right)}{2 B r^3 \left(r-\alpha  B e^{\phi _0} \phi _0'\right)}\nn\\
&+\frac{\phi _1 \left(2 \alpha  e^{\phi _0} \left((2 \Lambda +1) r B'+B \left(3 r B'+4 (B-1) r \phi _0'-4 B+4\right)\right)-r^2 \left(4 r B'+B \left(r \phi _0' \left(r \phi _0'+2\right)+4\right)-4\right)\right)}{4 B r^4 \left(r-\alpha  B e^{\phi _0} \phi _0'\right)}\nn\\
&+\frac{H_2' \left(\alpha  (3 B-1) e^{\phi _0} \phi _0'-2 r\right)}{2 r \left(r-\alpha  B e^{\phi _0} \phi _0'\right)}+\frac{\alpha  (B-1) e^{\phi _0} \phi _1''}{r^2 \left(\alpha  B e^{\phi _0} \phi _0'-r\right)}+K''=-\frac{8 \pi  r A_{{lm}}^{{(0)}}}{A B r-\alpha  A B^2 e^{\phi _0} \phi _0'}\,,\label{eq:ein_eq6}
\end{align}
where $H_0,\,H_1,\,H_2,\,K$ and $\phi_0$ are the polar parity perturbation functions defined in Sec.~\ref{subsec:pert}.

In the above equations, we defined $\Lambda=(l+2)(l-1)/2$. Note that, due to the Bianchi identities, not all of the above equations are independent. The dilaton field equation \eqref{eq:scalarphi} at first order gives
\begin{align}
&\frac{d^2\phi_1}{dr_*^2}+\left(\frac{B r A' \left(r \phi _0'+1\right)+A \left(r B' \left(r \phi _0'+1\right)+2 B r \left(r \phi _0''+2 \phi _0'\right)+4 (\Lambda +1)\right)}{2 r^2}-\omega ^2\right)\phi_1+\alpha  B^2 e^{\phi _0} A' K''\nn\\
&-\frac{\alpha  A (B-1) B e^{\phi _0} H_0''}{r}-\frac{H_0' \left(\alpha  e^{\phi _0} \left(2 (B-1) B A'+A (3 B-1) B'\right)+A B r^2 \phi _0'\right)}{2 r}-\frac{B H_2' \left(\alpha  (3 B-1) e^{\phi _0} A'+A r^2 \phi _0'\right)}{2 r}\nn\\
&+\alpha  H_2 e^{\phi _0} \left(B^2 r \left(A'\right)^2+A \left(-B \left(2 B r A''+A' \left(3 r B'+2 \Lambda +2\right)-2 r \omega ^2\right)-2 r \omega ^2\right)\right)+\frac{\alpha  A H_0 (\Lambda +1) e^{\phi _0} B'}{r^2}\nn\\
&+\frac{B K' \left(\alpha  e^{\phi _0} \left(A \left(2 B r A''+A' \left(3 r B'+4 B\right)\right)-B r \left(A'\right)^2\right)+2 A^2 r^2 \phi _0'\right)}{2 A r}-i H_1 \left(\frac{\alpha  (3 B-1) \omega  e^{\phi _0} B'}{r}+B r \omega  \phi _0'\right)\nn\\
&+\frac{\alpha  K e^{\phi _0} \left(A B' \left(2 r \omega ^2-\Lambda  A'\right)+B \Lambda  \left(\left(A'\right)^2-2 A A''\right)\right)}{2 A r}-\frac{2 i \alpha  (B-1) B \omega  e^{\phi _0} H_1'}{r}=0.
\label{eq:dilaton_per}
\end{align}
\end{widetext}
To obtain the system as \eqref{eq:perturbations} we do the following: we use Eqs.~\eqref{eq:ein_eq3}--\eqref{eq:ein_eq5} to eliminate $H_0$, $H_0'$, and $H_2$ in terms of $K$, $H_1$, $\phi_1$, and their first derivatives, as well as source terms, substituting them in Eqs.~\eqref{eq:ein_eq1}, \eqref{eq:ein_eq2} and \eqref{eq:dilaton_per}. In this way we can write
\be
\left(
\begin{array}{c}
 H_1' \\
 K' \\
 \phi _1' \\
 \phi _1''
\end{array}\right)
+\left(\begin{array}{c}
V_{11} \;\; V_{12} \;\; V_{13} \;\; V_{14} \\
V_{21}\;\; V_{22}\;\; V_{23} \;\; V_{24} \\
V_{31}\;\; V_{32}\;\; V_{33}\;\; V_{34} \\
V_{41}\;\; V_{42}\;\; V_{43} \;\;V_{44}
\end{array}\right)\left(
\begin{array}{c}
 H_1 \\
 K \\
 \phi _1\\
 \phi _1'
\end{array}\right)
=\left(
\begin{array}{c}
 S_1 \\
 S_2 \\
 S_3 \\
 S_4
\end{array}\right),
\label{eq:expanper}
\ee
which is the extended form of Eq.~\eqref{eq:perturbations} for $j={\rm p}$. Obviously, $V_{3k}=0$ for $k\neq 4$, $V_{34}=-1$, and $S_3=0$. Due to the complexity of the components of $\bm V_p$, we shall not show them here, but they can be seen in the supplementary Mathematica notebook \cite{ref:webpage}. 
The boundary conditions for the perturbations can be found with the aid of the expansions of the background metric and dilaton field. Note that, for QNMs, due to the natural divergence at spatial infinity, one must consider a high-order expansion of the perturbations at infinity.

For $l=0,1$, simpler gauge choices can be chosen~\cite{Zerilli:1971wd}. First, for $l=1$ we notice that we can pick a gauge in which $K$ vanishes identically. Therefore, we can use the first two equations in~\eqref{eq:expanper} to eliminate $H_1'$ and $H_1$ in favor of $\phi_1$, $\phi_1'$. For $l=0$, we can choose a gauge in which both $K$ and $H_1$ vanish and again the equations are reduced to a second-order equation for the dilaton perturbation. Note that this is possible because some harmonics in the expansion are identically zero in the $l=0$ and $l=1$ cases.

\subsection{Axial sector}
The equations for the axial sector are much simpler. The fundamental equations can be written as
\begin{align}
&h_1'-\frac{A A' B'+B \left(A \left(2 A''+A \phi '^2\right)-A'^2\right)}{2 A B \left(2 A-\alpha  B e^{\phi } A' \phi '\right)}r  h_1\nn\\
&+\frac{\alpha  e^{\phi } \left(B' \phi '+2 B \left(\phi ''+\phi '^2\right)\right)-2}{B \left(\alpha  B e^{\phi } A' \phi '-2 A\right)}i \omega  h_0=0,\label{eq:dh1}\\
&h_0'+\frac{r^2 \omega ^2	-\alpha  B e^{\phi } \phi ' \left(r \omega ^2-\Lambda  A'\right)-2 \Lambda  A}{r \omega  \left(r-\alpha  B e^{\phi } \phi '\right)}i h_1\nn\\
&-\frac{2 }{r}h_0=0\,,\label{eq:dh0}\\
&h_0''+i\omega h_1'+\frac{-2 A''-\frac{2 A'}{r}+\frac{A \left(2 B'+r B \phi '^2\right)}{B \left(\alpha  B e^{\phi } \phi '-r\right)}}{2 A'}h_0'\nn\\
&+\frac{i \omega  \left(-2 A''+\frac{2 A'}{r}+\frac{A \left(2 B'+r B \phi '^2\right)}{B \left(\alpha  B e^{\phi } \phi '-r\right)}\right)}{2 A'}h_1\nn\\
&+[{r A B \left(r-\alpha  B e^{\phi } \phi '\right)}]^{-1} \{B A' \left(r-\alpha  B e^{\phi } \phi '\right)\nn\\
&+A \left[\alpha  e^{\phi } \left((3 B+\Lambda ) B' \phi '+2 B (B+\Lambda ) \left(\phi ''+\phi '^2\right)\right)\right.\nn\\
&\left.-r B'-2 B-2 \Lambda \right]\}h_0=0.\label{eq:ddh0}
\end{align}
where $h_0,\,h_1$ are the axial parity perturbation functions defined in Sec.~\ref{subsec:pert}.

The above differential equations are already in the form of Eq.~\eqref{eq:perturbations}. As mentioned in the main text, the particle does not induce perturbation in the axial sector, and hence there are no source terms in the above equations.

For $l=0$, the tensorial harmonics multiplying the functions $h_0$ and $h_1$ vanish identically, and therefore the axial perturbation vanish identically.
 
For $l=1$, the analysis of the perturbations follows in a very similar manner as the one in GR~\cite{Zerilli:1971wd,Sago:2002fe}. One of the harmonics vanishes identically, and one is left with only two equations, namely~\eqref{eq:dh0} and \eqref{eq:ddh0}. We can exploit the gauge freedom to set either $h_1$ or $h_0$ to zero. The remaining equations, which are asymptotically GR, have a nonradiative behavior at infinity and contribute only to give an infinitesimal angular momentum to the BH.

\subsection{Source terms}

As mentioned in the main text, the particle stress-energy tensor can be expanded in spherical harmonics. The procedure to obtain the coefficients is outlined in Refs.~\cite{Zerilli:1971wd,Sago:2002fe,Martel:2005ir}. In the frequency domain, the source functions for a particle falling radially into the BH are given by
\begin{align}
A_{{lm}}^{{(0)}}&=\sqrt{l+\frac{1}{2}}\frac{\gamma ^2  e^{i \omega  T}}{2 \pi  r^2 \sqrt{\gamma ^2-{A}}},\\
A_{{lm}}^{{(1)}}&=-\sqrt{2 l+1}\frac{i \gamma   e^{i \omega  T}}{2 \pi  \sqrt{{A}} \sqrt{{B}} r^2},\\
A_{{lm}}&=\sqrt{l+\frac{1}{2}}\frac{ \sqrt{\gamma ^2-{A}} e^{i \omega  T}}{2 \pi  {A} {B} r^2},
\end{align}
where $\gamma(\geq 1)$ is the specific energy of the particle (boost parameter) and $T$ is the time trajectory of the particle, as a function of the radial coordinate. The function $T$ can be obtained by solving the differential equation
\be
T'=-\frac{\gamma }{\sqrt{A B} \sqrt{\gamma ^2-A}}.
\label{eq:time_geo}
\ee
We can solve this equation together with the perturbed equations, imposing that $T=0$ at the numerical horizon, without loss of generality. We note that in the small-$\zeta$ approximation for the metric Eq.~\eqref{eq:time_geo} can be solved analytically, expanding the right-hand side in powers of $\zeta$.

\bibliography{biblio}

\end{document}